\documentclass[journal,twocolumn]{IEEEtran}
\usepackage{cite}

\usepackage[none]{hyphenat}

\ifCLASSINFOpdf
   \usepackage[pdftex]{graphicx}
\else
   \usepackage[dvips]{graphicx}
\fi
\usepackage[cmex10]{amsmath}
\usepackage{amsfonts}
\usepackage{amssymb}
\usepackage{bm}
\newtheorem{theorem}{Theorem}
\newtheorem{lemma}{Lemma}
\newtheorem{proposition}{Proposition}

\begin{document}
%
\title{Cramer-Rao Lower Bounds for the Joint Delay-Doppler Estimation of an Extended Target}
%
%
%

\author{Tong Zhao,
\and
Tianyao Huang
\\
Radar Research (Beijing), Leihua Electronic Technology Institute, AVIC
}

\maketitle

\begin{abstract}
\boldmath
The problem on the Cramer-Rao Lower Bounds (CRLBs) for the joint time delay and Doppler stretch estimation of an extended target is considered in this paper. The integral representations of the CRLBs for both the time delay and the Doppler stretch are derived. To facilitate computation and analysis, series representations and approximations of the CRLBs are introduced. According to these series representations, the impact of several waveform parameters on the estimation accuracy is investigated, which reveals that the CRLB of the Doppler stretch is inversely proportional to the effective time-bandwidth product of the waveform. This conclusion generalizes a previous result in the narrowband case. The popular wideband ambiguity function (WBAF) based delay-Doppler estimator is evaluated and compared with the CRLBs through numerical experiments. Our results indicate that the WBAF estimator, originally derived from a single scatterer model, is not suitable for the parameter estimation of an extended target.
\end{abstract}



%
\IEEEpeerreviewmaketitle

\section{Introduction}
%
%
%
%
\IEEEPARstart{T}{he} joint estimation of time delay and Doppler stretch of a noise contaminated signal is a fundamental problem in radar and sonar systems,
and has been extensively addressed for the case involving narrowband signals \cite{M.Kay1993,923295,4217888,Johnson2008,Pourhomayoun2014,21705,260135,709535,1186885,4782982,5471174,6198368}.
In many modern sensing applications, however, wideband signals are utilized and the narrowband model may not be applicable in these situations.
A narrowband model is appropriate when $BT\ll c/2v$ \cite{Weiss1994}, where $B$ and $T$ are the bandwidth and the duration of the transmitted signal, respectively,
$v$ is the relative velocity between the target and the sensor, and $c$ is the propagation speed of the signal. In
imaging radars, for instance, signals with a large bandwidth $B$ are usually employed since high range resolution (HRR) is needed.
For systems requiring low interception probability (LPI), as another example, an effective approach is to emit a low-power signal with a
large $BT$ such that the energy is spread over a wider region in the time and/or frequency domain. In either case, the wideband model is more appropriate.
Meanwhile, the narrowband assumption may not be justifiable in some sonar systems \cite{Weiss1994}. The propagation velocity of sound in water is roughly $c = 1500 \mathrm{m/s}$.
If the relative velocity between the target and the sonar is $v = 22.5 \mathrm{m/s}$, it yields $c/2v=33.3 $. Thus, even a signal with $BT>10$ may not qualify as a narrowband signal.

There are significant differences in modeling echoes between wideband and narrowband signals. Firstly, a target can be modeled as a point scatterer under the narrowband
assumption. In contrast, a wideband signal entails a higher range resolution. Thus a target may span over several adjacent range units and should be described with multiple
scatterers\cite{Tang1996,710565,Liu2008a,5978228,6323291}. In this case, we refer to the target as an extended target. Secondly, the Doppler
effect on the echoes in the narrowband model is approximated by the shift of the carrier frequency while the complex
envelope of the transmitted waveform is assumed to be unaffected. For wideband signals, however, the Doppler effect on the
complex envelope must be taken into account \cite{Weiss1994}.

As a lower bound for the variance of any unbiased estimator,
the Cramer-Rao Lower Bound (CRLB) is an important tool for performance evaluation of various estimation methods\cite{M.Kay1993,Wei2007,5656265,6624548,Jin1995,1513159,6409764}.
Due to the asymptotically efficient property of the maximum likelihood estimator (MLE),
the CRLB can be used to predict the performance of the MLE. In addition, the CRLB has been employed
as a criterion for optimal waveform selections \cite{Jin1995,1513159,6409764,Huang2014}.

The CRLB for the joint time delay and Doppler shift estimation with narrowband signals has been investigated in numerous studies (e.g. \cite{M.Kay1993,Johnson2008,Pourhomayoun2014,923295,4217888} and
references therein), while the counterpart for wideband signals has received less attention. Specifically, \cite{Wei2007,Jin1995} concern the CRLB for a general wideband signal along with a single scatterer model.
Exploiting an extended target model,  \cite{Liu2014} consider the CRLBs of the velocity and HRR profiles estimation with a step frequency signal. The study in  \cite{5464895} examines  the CRLB for a noncoherent
multiple-input multiple-output (MIMO) radar system in which the signals transmitted from different transmitters are assumed to be narrowband.
In this paper, we consider more general wideband sensing systems and derive the CRLB for an arbitrary wideband signal along with an extended target model.

It is shown in \cite{M.Kay1993, 6409764,5656265} that under the wideband model for a single scatterer, the CRLB of the time delay estimation is inversely proportional to the
effective bandwidth of the transmitted signal. Under a similar condition, \cite{Jin1995} proves that increasing the effective time-bandwidth product can improve the joint
estimation accuracy of the time delay and the Doppler stretch. Nevertheless, how the waveform parameters affect the CRLB of the time delay as well as the CRLB of the Doppler stretch are not considered separately.
\cite{1683396} discusses the effect of the bandwidth on the range estimation accuracy in a multipath environment through simulation and shows that ranging error diminishes with an increasing bandwidth.
In this paper, we take an analytical approach and study the effects of waveform parameters on the CRLB of both the time delay and the Doppler stretch for the wideband model along with an extended
target. As shown in this paper, the CRLB of the time delay is inversely proportional to the effective bandwidth, and the CRLB of the Doppler stretch is inversely proportional to the
effective time-bandwidth product. These conclusions for the wideband model are a generalization of the counterpart in the narrowband case.

The rest of this paper is organized as follows.
Section \ref{sec:a4} establishes the signal model and defines the waveform parameters under investigation. In Section \ref{sec:a5},
the CRLBs of the time delay and the Doppler stretch are derived and discussed. The integral representations of the CRLBs are presented firstly and, for the convenience of analysis,
series representations and approximations are presented. Based on the series representations, the influences of the effective bandwidth and the effective duration on the CRLBs are analyzed.
Section \ref{sec:a6} provides some numerical examples, in which the performance of the wideband ambiguity function (WBAF) based delay-Doppler estimator is evaluated. Section \ref{sec:a7} contains the conclusions.




\section{Modeling and problem statement}{\label{sec:a4}}

Consider an extended target which contains $P$ ideal scatterers and is moving along the line of sight (LOS) with
a constant radial velocity $v$ relative to the sensor. The
velocity is positive if the target is moving away from the
sensor. We assume that $\tau_p=\tau+(p-1)\Delta$ is the time delay of the $p$th scatterer, where $\Delta$ is the sampling interval. Thus, the scatterers are equally spaced along the LOS. $\gamma=(c-v)/(c+v)$ is the Doppler stretch , where $c$ is the propagation speed of the transmitted waves.
$L =\frac{1}{2}c(\tau_P-\tau_1)$ is the size of the target.
Let $s(t)$ be the complex envelope of the transmitted signal which is time-limited to $\left[0,T\right]$,
that is, $s(t)=0$ if $t \notin[0,T]$. Thus, the complex envelope of the echo can be modeled as
\begin{equation}{\label{equ:a22}}
y(t)=\sum\limits_{p=1}^P x_p s(\gamma(t-\tau_p))+w(t),
\end{equation}
where $x_p$ accounts for the propagation attenuation
and the influence of the Doppler stretch on the signal energy. The noise $w(t)$ is considered as a bandlimited complex Gaussian random process, where $\mathrm{Re}\left\{w(t)\right\}$ and $\mathrm{Im}\left\{w(t)\right\}$ are mutually independent with a bandwidth $1/(2\Delta)$ and power spectral density
$N_0/2$.
Sample the echo at the rate of $1/\Delta$, and (\ref{equ:a22}) turns into
\begin{equation}\label{equ:a23}
y_n=\sum\limits_{p=1}^P x_p  s(\gamma(n\Delta-\tau_p))+w_n,
n=0,1,...,N-1,
\end{equation}
where $y_n=y(n\Delta)$, the noise $w_n=w(n\Delta)$ is distributed as $\mathbb{C} N(0,\sigma^2) $ and
$\sigma^2\Delta=N_0$ \cite{M.Kay1993}. Rewrite (\ref{equ:a23}) as
\begin{equation}{\label{equ:a76}}
{\bf y}={\bf\Phi} {\bf x}+{\bf w},
\end{equation}
where ${\bf y}=\left[y_0,...,y_{N-1}\right]^T \in \mathbb{C}^{N \times 1} $ is
the observation vector, ${\bf\Phi} \in \mathbb{C} ^{N\times P} $ is the
complex measurement matrix with $\Phi_{ij}=s(\gamma((i-1)\Delta-\tau_j))$. Scattering coefficients ${\bf x}=\left[x_1,...,x_P\right]^T \in \mathbb{C}^{P\times 1} $
represent the
high-range-resolution profile of the target. The noise vector ${\bf
w}=\left[w_0,...,w_{N-1}\right]^T \in \mathbb{C}^{N\times 1}$ is distributed as $
\mathbb{C}N({\bf 0},\sigma^2{\bf I}) $. The parameters under estimation are $\bm{\theta}=\left[\tau,\gamma,{\bf a}^T,{\bf b}^T \right] ^T$, where ${\bf a}=\mathrm{Re}\{{\bf x}\}$ and ${\bf b}=\mathrm{Im}\{{\bf x}\}$. In addition, following assumptions are made:

\emph{Assumption 1}: For $p=1,...,P$, we have
$\gamma((N-1)\Delta-\tau_p) \ge T$. It indicates that the echo  are completely sampled.

\emph{Assumption 2}: Both $\frac{\tau_p}{\Delta}$ and
$\frac{T}{\gamma \Delta} $ are considered as integers. It suggests
that the sampling interval $\Delta$ is small enough.

\emph{Assumption 3}: $s(t)$ has derivatives of all orders throughout
$(-\infty,+\infty) $ and there exist constants $C_1,C_2 > 0 $ such that
\begin{equation}{\label{equ:a89}}
\left|M_i^{(k)}\right|<C_1e^{kC_2},\left|\widetilde{M}_i^{(k)}\right|<C_1e^{kC_2},i=0,1,2,k\in \mathbb{N},
\end{equation}
where
\begin{equation}
M_i^{(k)}= \int_{-\infty}^{+\infty}  t^i \left|s^{(k)}(t)\right|^2dt,
\end{equation}
\begin{equation}
\widetilde{M}_i^{(k)}=\mathrm{Im} \left\{ \int_{-\infty}^{+\infty}  t^i s^{*(k)}(t)s^{(k+1)}(t)dt\right\}.
\end{equation}
The notation $s^{(m)}(t)$ is the abbreviation of $\frac{d^{m}s}{dt^{m}}(t)$. The regularity condition (\ref{equ:a89}) ensures the interchangeability between integrals and limits in Subsection \ref{sec:a10}.

The parameter $M_0^{(0)}$ is the energy of the transmitted waveform, while $M_0^{(1)}$ and $M_2^{(1)}$ can be considered as a measure of the bandwidth and the duration, respectively. The {\it effective bandwidth} of $s(t)$ is defined by \cite{Rihaczek1985}
\begin{equation}
\bar{B}=\left(\frac{M_0^{(1)}}{M_0^{(0)}}\right)^{\frac{1}{2}}=\left(\frac{\int\limits_{-\infty}^{+\infty} \left|s^{(1)}(t)\right|^2dt}{\int\limits_{-\infty}^{+\infty} \left|s(t)\right|^2dt}\right)^{\frac{1}{2}}.
\end{equation}
According to the properties of the Fourier transformation, $\bar{B}$ measures the spread of the signal $s(t)$ in the frequency domain in a root mean square (RMS) sense and thus we also refer to it as the RMS bandwidth. We define the {\it effective duration} $\bar{T}$ as
\begin{equation}{\label{equ:a88}}
\bar{T}=\left(\frac{M_2^{(1)}}{M_0^{(1)}}\right)^{\frac{1}{2}}=\left(\frac{\int\limits_{-\infty}^{+\infty} \left|s^{(1)}(t)\right|^2t^2dt}{\int\limits_{-\infty}^{+\infty} \left|s^{(1)}(t)\right|^2dt}\right)^{\frac{1}{2}}.
\end{equation}
and refer to $\bar{B}\bar{T}$ as the {\it effective time-bandwidth product}. Note that the definition of the effective duration in this paper is different from that in \cite{Rihaczek1985}, where the effective duration is defined by
\begin{equation}
T_2=\left(\frac{M_2^{(0)}}{M_0^{(0)}}\right)^{\frac{1}{2}}=\left(\frac{\int\limits_{-\infty}^{+\infty} \left|s(t)\right|^2t^2dt}{\int\limits_{-\infty}^{+\infty} \left|s(t)\right|^2dt}\right)^{\frac{1}{2}}.
\end{equation}
For a narrowband signal
\begin{equation}{\label{equ:a77}}
s(t)=\mathrm{exp}\left\{ j2\pi f_c t\right\}(u(t)-u(t-T)),
\end{equation}
where $u(t)$ is the unit step function, we have $\bar{B}=f_c$ and $\bar{T}=T_2=\frac{\sqrt{3}}{3}T$. Thus, these two definitions on the effective duration are in accord for narrowband signals. As shown in the Subsection \ref{sec:a9} and  \ref{sec:a8}, the CRLBs of the time delay and the Doppler stretch for a wideband signal are largely influenced by $\bar{T}$, not $T_2$. Therefore, we will henceforth use the definition of the effective duration (\ref{equ:a88}) in the following sections.

Finally, some matrices and notations are
introduced. We define ${\bf \Gamma}^{(k)}=\left[\Gamma^{(k)}_{ij}\right]\in  \mathbb{R}^{P\times P}$,
$ 1 \le i,j \le P, k=0,1,2,...$ , where
\begin{equation}
\Gamma^{(k)}_{ij} =(\tau_i-\tau_j)^k=\Delta_{ij}^k.
\end{equation}
Particularly, ${\bf \Gamma}^{(0)}  ={\bf 1}_{P \times P}={\bf 1}{\bf 1}^T$, that is, all elements in ${\bf \Gamma}^{(0)} $ are equal to $1$.  In addition, the notation $f(x)=O(g(x))$, as $x\rightarrow x_0$ means that there exist constants $C_3, C_4 > 0$, such that
\begin{equation}{\label{equ:a102}}
 \limsup_{x\to x_0}\left|\frac{f(x)}{g(x)}\right|=\lim\limits_{\delta \to 0}\sup\limits_{x \in (x_0-\delta,x_0+\delta)}\left|\frac{f(x)}{g(x)}\right|<C_3,
\end{equation}.
\begin{equation}
\liminf_{x\to x_0}\left|\frac{f(x)}{g(x)}\right|=\lim\limits_{\delta \to 0}\inf\limits_{x \in (x_0-\delta,x_0+\delta)}\left|\frac{f(x)}{g(x)}\right|>C_4,
\end{equation}.

\section{Derivations of the CRLBs}{\label{sec:a5}}
\subsection{Integral representations of CRLBs}
The CRLB is a lower bound for the variance of any unbiased estimator and is usually used as a benchmark to evaluate the performance of estimators.  The parameters under estimation are $\bm{\theta}=\left[\tau,\gamma,{\bf a}^T,{\bf b}^T \right] ^T$. According to \cite{M.Kay1993}, the covariance matrix of any unbiased
estimator $\hat{\bm{\theta}}$ satisfies
\begin{equation}
{\bf C}_{\hat{\bm{\theta}}} \triangleq
E\left\{(\hat{\bm{\theta}}-\bm{\theta})\cdot(\hat{\bm{\theta}}-\bm{\theta})^T\right\}\ge
{\bf FIM}^{-1} ,
\end{equation}
where ${\bf FIM} \in \mathbb{R}^{(2P+2)\times (2P+2)}$ is the Fisher information matrix defined by
\begin{equation}
{\bf FIM}=
E\left\{\bm{ \triangledown}_{\bm{\theta}}(\mathrm{ln}p({\bf y};\bm{\theta})) \cdot
\bm{ \triangledown}_{\bm{\theta}}^T(\mathrm{ln}p({\bf y};\bm{\theta}))\right\}
\end{equation}
with
\begin{equation}
\bm{\triangledown}_{\bm{\theta}} \triangleq\left[\frac{\partial}{\partial \tau},
\frac{\partial}{\partial\gamma} , \frac{\partial}{\partial x_1}
,...,\frac{\partial}{\partial x_{P}}\right]^T.
\end{equation}
The matrix inequality ${\bf A} \geq {\bf B}$ means ${\bf A}-{\bf B}$ is positive semidefinite. The observation vector ${\bf y}$ in (\ref{equ:a76}) is distributed as $\mathbb{C}N ( \bm{\mu}(\bm{\theta}),\sigma^2{\bf I})$ with $\bm{\mu}(\bm{\theta})= {\bf \Phi}{\bf x}$, and thus the Fisher information matrix can be calculated by \cite{M.Kay1993}
\begin{equation}{\label{equ:a33}}
\left[{\bf FIM}\right]_{ij}=\frac{2}{\sigma^2} \mathrm{Re}\left\{\frac{\partial
\bm{\mu}^H(\bm{\theta})}{\partial \theta_i}\frac{\partial
\bm{\mu}(\bm{\theta})}{\partial \theta_j}\right\}.
\end{equation}
The CRLB of $\bm{\theta}$ is given by the diagonal elements of ${\bf FIM}^{-1}$.
Partition {\bf FIM} as
\begin{equation}
 {\bf FIM}= \left[
\begin{array}{cccc}
F_{11}& F_{12}& {\bf F}_{31}^T & {\bf F}_{41}^T\\
F_{12}& F_{22}&{\bf F}_{32}^T & {\bf F}_{42}^T\\
{{\bf F}_{31}}&{\bf F}_{32}&{\bf F}_{33}&-{\bf F}_{43}\\
{{\bf F}_{41}}&{\bf F}_{42}&{\bf F}_{43}&{\bf F}_{33}\\
\end{array}
\right],
 \end{equation}
 where ${\bf F}_{ij} \in \mathbb{R}^{P_i \times P_j}$ with $P_1=P_2=1$ and $P_3=P_4=P$. The elements of ${\bf FIM}$ are calculated by
\begin{align}
&F_{11}=\frac{2}{\sigma^2}\mathrm{Re} \left\{ {\bf x}^H \frac{\partial \Phi^H}{\partial \tau}\frac{\partial \Phi}{\partial \tau}{\bf x} \right\},{\bf F}_{31}=\frac{2}{\sigma^2} \mathrm{Re}\left\{\Phi^H\frac{\partial \Phi}{\partial \tau} {\bf x}\right \}, \nonumber\\
&F_{12}= \frac{2}{\sigma^2}\mathrm{Re}\left\{ {\bf x}^H\frac{\partial \Phi^H}{\partial \tau}\frac{\partial \Phi}{\partial \gamma}{\bf x} \right\},{\bf F}_{32}=\frac{2}{\sigma^2}\mathrm{Re}\left\{  \Phi ^H\frac{\partial \Phi}{\partial \gamma}{\bf x}\right\}, \nonumber \\
&F_{22}= \frac{2}{\sigma^2}\mathrm{Re}\left\{{\bf x}^H \frac{\partial \Phi^H}{\partial \gamma}\frac{\partial \Phi}{\partial \gamma}{\bf x}\right \},{\bf F}_{33}= \frac{2}{\sigma^2}\mathrm{Re}\left\{\Phi^H\Phi\right\}, \nonumber\\
&{\bf F}_{41}=\frac{2}{\sigma^2} \mathrm{Im}\left\{\Phi^H\frac{\partial \Phi}{\partial \tau} {\bf x}\right \},{\bf F}_{42}=\frac{2}{\sigma^2} \mathrm{Im}\left\{\Phi^H\frac{\partial \Phi}{\partial \gamma} {\bf x}\right \} ,\nonumber\\
&{\bf F}_{43}=\frac{2}{\sigma^2} \mathrm{Im}\left\{\Phi^H\Phi\right\}.
\end{align}
 The CRLB of the time delay and Doppler stretch are given by
\begin{equation}{\label{equ:a24}}
\mathrm{var}(\tau) \ge \mathrm{CRLB}_{\tau}=a_{22}/\left(a_{11}a_{22}-a_{12}^2\right) ,
\end{equation}
\begin{equation}{\label{equ:a25}}
\mathrm{var}(\gamma) \ge \mathrm{CRLB}_{\gamma}=a_{11}/\left(a_{11}a_{22}-a_{12}^2\right),
\end{equation}
where
\begin{equation}
a_{ij}=F_{ij}-\left[\begin{array}{cc}{\bf F}_{3i}^T &{\bf F}_{4i}^T\end{array}\right] \left[\begin{array}{cc} {\bf F}_{33} &-{\bf F}_{43}\\{\bf F}_{43} &{\bf F}_{33} \end{array}\right] ^{-1}
\left[\begin{array}{c} {\bf F}_{3j}\\{\bf F}_{4j}\end{array}\right].
\end{equation}
Define
 \begin{equation}{\label{equ:a13}}
\overline{{\bf F}}_{31}={\bf F}_{31}+j{\bf F}_{41}=\frac{2}{\sigma^2}\Phi^H\frac{\partial \Phi}{\partial \tau} {\bf x} ,
\end{equation}
\begin{equation}
\overline{{\bf F}}_{32}={\bf F}_{32}+j{\bf F}_{42}=\frac{2}{\sigma^2}\Phi ^H\frac{\partial \Phi}{\partial \gamma}{\bf x} ,
\end{equation}
\begin{equation}{\label{equ:a107}}
\overline{{\bf F}}_{33}={\bf F}_{33}+j{\bf F}_{43}=\frac{2}{\sigma^2}\Phi^H\Phi .
\end{equation}
Then, we have
\begin{equation}{\label{equ:a87}}
a_{ij}=F_{ij}-\mathrm{Re}\left\{\overline{{\bf F}}_{3i}^H\overline{{\bf F}}_{33}^{-1}\overline{{\bf F}}_{3j}\right\},
\end{equation}
Note that $\overline{{\bf F}}_{33}^{-1}$ exists due to the positive definite property of ${\bf FIM}$. The elements of ${\bf FIM}$ are calculated as following
\begin{align}{\label{equ:a37}}
&F_{11}=\frac{2}{\sigma^2}\mathrm{Re} \left\{ {\bf x}^H \frac{\partial \Phi^H}{\partial \tau}\frac{\partial \Phi}{\partial \tau}{\bf x} \right\}=  \frac{2}{\sigma^2}\sum\limits_{i=1}^{P} \sum\limits_{j=1}^{P}  \sum\limits_{n=0}^{N-1} \nonumber\\
&\ \ \ \ \mathrm{Re} \left\{   x_i^{*}x_j \frac{ \partial s^{*}( \gamma(n\Delta-\tau_i))}{\partial \tau}\frac{\partial s(\gamma(n\Delta-\tau_j))}{\partial \tau} \right\} \nonumber\\
&\approx  \frac{2\gamma}{N_0}  \sum\limits_{i=1}^{P}
\sum\limits_{j=1}^{P}  x_i^*x_j
\int_{-\infty}^{+\infty} s^{*(1)}(t)
s^{(1)}(t+\gamma\Delta_{ij})dt,
\end{align}
In the last line, the summation is approximated with an integral by letting $\Delta \rightarrow 0$, which is based on the assumption that the sampling interval $\Delta$ is small enough. Similarly, we have
\begin{align}{\label{equ:a26}}
F_{12}&= \frac{2}{\sigma^2}\mathrm{Re}\left\{ {\bf x}^H\frac{\partial \Phi^H}{\partial \tau}\frac{\partial \Phi}{\partial \gamma}{\bf x} \right\}\approx -\frac{2}{\gamma N_0  }
\sum\limits_{i=1}^{P}\sum\limits_{j=1}^{P}  \nonumber\\
& \mathrm{Re}\left\{x_i^{*}x_j
\int_{-\infty}^{+\infty}t
s^{*(1)}(t+\gamma\Delta_{ji}) s^{(1)}(t)  dt\right\} ,
\end{align}
\begin{align}
&F_{22}= \frac{2}{\sigma^2}\mathrm{Re}\left\{{\bf x}^H \frac{\partial \Phi^H}{\partial \gamma}\frac{\partial \Phi}{\partial \gamma}{\bf x}\right \} \approx \frac{2}{\gamma^3 N_0 } \sum\limits_{i=1}^P\sum\limits_{j=1}^{P} \nonumber\\
&x_i^{*}x_j\int_{-\infty}^{+\infty}  t ( t+\gamma\Delta_{ij}) s^{*(1)}(t) s^{(1)}(t+\gamma\Delta_{ij}) dt,
\end{align}
\begin{align}
\left[\overline{{\bf F}}_{31}\right]_{i1} \approx - \frac{2}{N_0 } \sum\limits_{j=1}^P
x_j\int_{-\infty}^{+\infty} s^{*}(t)
s^{(1)}(t+\gamma\Delta_{ij}) dt,
\end{align}
\begin{align}
\left[\overline{{\bf F}}_{32}\right]_{i1}\approx \frac{2}{\gamma^2 N_0 } \sum\limits_{j=1}^P x_j
\int_{-\infty}^{+\infty} t
s^{*}(t+\gamma\Delta_{ji})s^{(1)}(t) dt,
\end{align}
\begin{align}{\label{equ:a27}}
\left[\overline{{\bf F}}_{33}\right]_{ij}\approx\frac{2}{\gamma N_0 } \int_{-\infty}^{+\infty} s^{*}(t) s(t+\gamma\Delta_{ij})dt.
\end{align}
Substituting (\ref{equ:a37})-(\ref{equ:a27}) into (\ref{equ:a24})-(\ref{equ:a25}) gives the CRLBs.
\subsection{Series representations of CRLBs}{\label{sec:a10}}
The previous representations have a limitation that they are not helpful to analyze the properties of the CRLBs. One reason is that (\ref{equ:a37})-(\ref{equ:a27}) invlove many complicated integrals of which the integrands depend on both the waveform and the target.
To address this issue, we replace functions $s\left(t+\gamma\Delta_{ij}\right)$ and $s^{(1)}\left(t+\gamma \Delta_{ij}\right)$ in (\ref{equ:a37})-(\ref{equ:a27}) with their Taylor series, respectively, and then rewrite the CRLBs in the form of series. As shown in (\ref{equ:a39})-(\ref{equ:a29}), these series representations only consist of integrals $M_i^{(k)}, \widetilde{M}_i^{(k)}, i=1,2,3, k \in \mathbb{N}$,
which are relatively uncomplicated and only depend on waveform. Notice that some important waveform parameters, such as the effective bandwidth and the effective duration, are directly determined by these integrals.
Therefore, it is easier to analyze the CRLBs by employing the series representations.
In this subsection, the series representations of CRLBs are derived and then some approximations on CRLBs are presented.

Based on the Taylor series \cite{Zorich2004I}, we have
\begin{equation}{\label{equ:a51}}
s\left(t+\gamma\Delta_{ij}\right) =
\sum\limits_{m=0}^{+\infty}\frac{\gamma^{m}\Delta_{ij}^{m}}{m!}s^{(m)}(t), \forall t \in \mathbb{R},
\end{equation}
\begin{equation}{\label{equ:a52}}
 s^{(1)}\left(t+\gamma\Delta_{ij}\right) =
\sum\limits_{m=1}^{+\infty}\frac{\gamma ^{m-1}\Delta_{ij}^{m-1}}{(m-1)!}s^{(m)}(t), \forall t \in \mathbb{R},
\end{equation}
where $0!\triangleq 1 $. Substituting (\ref{equ:a51})-(\ref{equ:a52}) into (\ref{equ:a37})-(\ref{equ:a27}), applying Theorem \ref{theorem:12} (Appendix \ref{app:05}) and the Lebesgue's Dominated Convergence Theorem, which is one of the rules on the interchangeability between integral and limit \cite{H.L.Royden1988}, we obtain the CRLBs in the form of series, which are presented in the Appendix \ref{app:a13} due to their complicate expressions.

By (\ref{equ:a28})-(\ref{equ:a29}), the CRLBs can be approximated as
\begin{equation}{\label{equ:a30}}
\mathrm{CRLB}^{(K)}_{\tau}=a_{22}^{(K)}/  \left(a
_{11}^{(K)}a_{22}^{(K)}-\left(a_{12}^{(K)}\right)^2\right) ,
\end{equation}
\begin{equation}
\mathrm{CRLB}^{(K)}_{\gamma}=a_{11}^{(K)}/\left(a
_{11}^{(K)}a_{22}^{(K)}-\left(a_{12}^{(K)}\right)^2\right),
\end{equation}
where
\begin{equation}{\label{equ:a21}}
a_{ij}^{(K)}=F_{ij}^{(K)}-\mathrm{Re}\left\{\left(\overline{{\bf F}}_{3i}^{(K)}\right)^H\left(\overline{{\bf F}}_{33}^{(K)}\right)^{-1}\overline{{\bf F}}_{3j}^{(K)}\right\}.
\end{equation}
It is recommended to substitute ${\bf F}_{33}^{(K)}$ with its original value to avoid the possible singularity of ${\bf F}_{33}^{(K)}$. The next result provides an bound on the approximation error due to the truncation.
\begin{proposition}{\label{theorem:13}}
\begin{equation}{\label{equ:a83}}
\mathrm{CRLB}^{(K)}_{\tau}-\mathrm{CRLB}_{\tau}=O\left(\frac{ \left( 2L \gamma \exp\{C_2\}/ c\right)^{K+1}}{(K+1)!}\right),
\end{equation}
\begin{equation}{\label{equ:a84}}
\mathrm{CRLB}^{(K)}_{\gamma}-\mathrm{CRLB}_{\gamma}=O\left(\frac{ \left( 2L \gamma \exp\{C_2\}/ c\right)^{K+1}}{(K+1)!}\right),
\end{equation}
as $K \to +\infty$, where $C_2$ is the constant in (\ref{equ:a89}).
\end{proposition}
\begin{IEEEproof}
See the Appendix \ref{app:a15}.
\end{IEEEproof}

Proposition \ref{theorem:13} indicates that the error of the approximate CRLBs is bounded, and thus the factors impacting on the error can be analyzed. According to (\ref{equ:a83})-(\ref{equ:a84}), a larger $K$ is required as the size of the target $L$ increases.

Finally, we consider a special case where the scattering coefficients $\bf x$ are real numbers. The CRLBs are given by (\ref{equ:a24})-(\ref{equ:a25}), where
\begin{equation}{\label{equ:a106}}
a_{ij}=F_{ij}-{\bf F}_{3i}^T{\bf F}_{33}^{-1}{\bf F}_{3j} ,i,j=1,2.
\end{equation}
Comparing (\ref{equ:a13})-(\ref{equ:a107}) with (\ref{equ:a32}), we have
\begin{equation}
{\bf F}_{3i}=\lim_{K\to +\infty}{\bf F}_{13i}^{(K)},i=1,2,3.
\end{equation}
Note that $F_{2ij}^{(K)}=0,i,j=1,2$, and thus the CRLBs do not rely on $\widetilde{M}_i^{(k)}$.  In Subsection \ref{sec:a9} and \ref{sec:a8} when we discuss the influences of waveform parameters on the CRLBs, it is reasonable to believe that $M_i^{(k)}$ and $\widetilde{M}_i^{(k)}/M_i^{(k)}$ are independent because $\widetilde{M}_i^{(k)}$ is a measurement of the correlation between $s^{(k)}$ and $s^{(k+1)}$. Furthermore, the interesting waveform parameters are defined by $M_i^{(k)}$ and the representations of the CRLBs are briefer in this case. Thus, we henceforth only consider the case of real-valued scattering coefficients without loss of generality and our development of Subsection \ref{sec:a9} and \ref{sec:a8} can be easily generalized to the complex case.
\subsection{Discussions on waveform parameters with $P=1$}{\label{sec:a9}}
In  previous subsections, the CRLBs for an extended target are derived.
When $P = 1$, the extended target becomes a single scatterer and ${\bf x}$ reduces to a scalar $x$.
Let $P=1$ in (\ref{equ:a24})-(\ref{equ:a25}), and the CRLBs of a single scatterer target are
\begin{equation}{\label{equ:a56}}
\mathrm{var}_{P=1}(\tau) \ge \mathrm{CRLB}_{\tau,P=1}=\frac{\tilde{a}_{22}}{\tilde{a}_{11}\tilde{a}_{22}-\tilde{a}_{12}^2 } ,
\end{equation}
\begin{equation}{\label{equ:a69}}
\mathrm{var}_{P=1}(\gamma) \ge \mathrm{CRLB}_{\gamma,P=1}=\frac{\tilde{a}_{11}}{\tilde{a}_{11}\tilde{a}_{22}-\tilde{a}_{12}^2},
\end{equation}
where
\begin{equation}{\label{equ:a62}}
\tilde{a}_{11}=\frac{2\gamma x^2M_0^{(1)}}{N_0}, \tilde{a}_{12}=-\frac{2 x^2M_1^{(1)}}{\gamma N_0},
\end{equation}
\begin{equation}{\label{equ:a57}}
\tilde{a}_{22}=\frac{2x^2}{\gamma^3 N_0}\left(M_2^{(1)}-\frac{M_0^{(0)}}{4}\right).
\end{equation}

We next investigate the influences of the effective bandwidth $\bar{B}$ and the effective duration $\bar{T}$ of the transmitted signal on the CRLBs. We assume that $\bar{B}$ and $\bar{T}$ are independent of $M_0^{(0)}$, which holds if an alteration of $M_0^{(0)}$ results from changing the amplitude of the transmitted signal.
\begin{theorem}{\label{theorem:a13}}
Let $M_0\le +\infty$,  $B_0\le +\infty$ and $T_0<+\infty$. Assume that 1) $\bar{B}$ and $\bar{T}$ are independent of $M_0^{(0)}$, 2) there exists a constant $\epsilon>0$ such that
\begin{equation}{\label{equ:a108}}
\lim\limits_{ \left(M_0^{(0)}, \bar{B}, \bar{T}\right) \rightarrow \left(M_0,B_0,T_0\right) }m\left(\left\{t \left|\left|s^{(1)}(t)\right|>\epsilon \right.\right\}\right)>0,
\end{equation}
where $m(A)$ is the Lebesgue measure of a set $A\subset \mathbb{R}$. Then, we have
\begin{align}{\label{equ:a64}}
\mathrm{CRLB}_{\tau, P=1}=O\left( \left(M_0^{(0)}\right)^{-1} \bar{B}^{-2}\right),
\end{align}
\begin{align}{\label{equ:a65}}
\mathrm{CRLB}_{\gamma, P=1}=O\left(\left(M_0^{(0)}\right)^{-1}\bar{B}^{-2}\bar{T}^{-2}\right),
\end{align}
as $\left(M_0^{(0)}, \bar{B}, \bar{T}\right) \rightarrow \left(M_0,B_0,T_0\right)$.
\end{theorem}
\begin{IEEEproof}
See the Appendix \ref{app:a12}.
\end{IEEEproof}
Notice that (\ref{equ:a108}) can be easily met in practice. Therefore, we conclude that 1) there exists a positive correlation between the effective bandwidth $\bar{B}$ and the estimation accuracy of the time delay, 2) there exists a positive correlation between the effective time-bandwidth product $\bar{B}\bar{T}$ and the estimation accuracy of the Doppler stretch.
\subsection{Discussions on waveform parameters in the general case}{\label{sec:a8}}
In this subsection, discussions about the influences of waveform parameters on CRLBs in the case of a single scatterer are generalized to the extended target situation. It is worth mentioning that an alteration of the effective bandwidth or the effective duration results in changes of $M_i^{(k)}$, $i=0,1,2,k \ge 2$, which also affect the CRLBs. Therefore, $\bar{B}$ and $\bar{T}$ influence the CRLBs partly through these waveform parameters. Notice that the leading terms in (\ref{equ:a39})-(\ref{equ:a29}) only contain $M_0^{(0)}$, $M_i^{(1)}$, $i=1,2,3$ and have no immediate relations with $M_i^{(k)}$, $k \ge 2$. Thus, it is believed that for an extended target, the energy, effective bandwidth and effective duration influence the CRLBs mainly through $M_i^{(k)}, k\le 1$ rather than $M_i^{(k)}$ or $M_i^{(k)}/M_i^{(1)}$, $k \ge 2$. To simplify the discussion, we assume that $M_0^{(0)}$, $\bar{B}$ and $\bar{T}$  influence the CRLBs through $M_i^{(k)}, k\le 1$.
\begin{theorem}\label{theorem:11}
Suppose that 1) $M_i^{(k)}/M_0^{(0)} $  are independent of $M_0^{(0)}$, $i=0,1,2, k \in \mathbb{N}^+$, 2) $M_0^{(0)}$, $\bar{B}$ and $\bar{T}$  influence the CRLBs through $M_i^{(k)}, k\le 1$, and 3) $\bar{T}$ and $\bar{B}$ are mutually independent. Then, for $M_0<+\infty$,  $B_0<+\infty$ and $T_0<+\infty$, we have
\begin{equation}{\label{equ:a73}}
\mathrm{CRLB}_{\tau}\approx O\left(\left(M_0^{(0)}\right)^{-1}\bar{B}^{-2} \right),
\end{equation}
\begin{equation}{\label{equ:a74}}
\mathrm{CRLB}_{\gamma}\approx O\left(\left(M_0^{(0)}\right)^{-1}\bar{B}^{-2}\bar{T}^{-2} \right),
\end{equation}
as $\left(M_0^{(0)}, \bar{B}, \bar{T}\right) \rightarrow \left(M_0,B_0,T_0\right)$.
\end{theorem}
\begin{IEEEproof}
See the Appendix \ref{app:a12}.
\end{IEEEproof}
Therefore, we concluded that 1) there exists a positive correlation between the estimation accuracy of the time delay and the effective bandwidth, 2) the estimation accuracy of the Doppler stretch is positive correlated to the effective time-bandwidth product.

Consider the narrowband signal (\ref{equ:a77}), we have
\begin{equation}
\mathrm{CRLB}_{\tau}=O\left(\left(M_0^{(0)}\right)^{-1}f_c^{-2} \right),
\end{equation}
\begin{equation}
\mathrm{CRLB}_{\gamma}=O\left(\left(M_0^{(0)}\right)^{-1}f_c^{-2}T^{-2} \right).
\end{equation}
The Doppler shift is defined by $f_d=\gamma f_c-f_c$. According to \cite{M.Kay1993}, the CRLB of the Doppler shift is given by
\begin{equation}
\mathrm{CRLB}_{f_d}=f_c^2\mathrm{CRLB}_{\gamma}=O\left(\left(M_0^{(0)}\right)^{-1}T^{-2} \right).
\end{equation}
It indicates that for narrowband signals, there exists a positive correlation between the estimation accuracy of the Doppler shift and the duration.
\section{Numerical Results}{\label{sec:a6}}
In this section, we compare the performances of several estimators with the derived CRLBs and provide numerical examples to illustrate the properties of CRLBs.

In the case where a narrowband signal is transmitted, a standard method to estimate the time delay and the Doppler stretch is to use the ambiguity function (AF) \cite{M.Kay1993,Rihaczek1985}, which is asymptotically efficient, that is, the estimator is unbiased and reaches the CRLB when the number of independent observations approaches to infinity \cite{Kendall1979}. For a wideband model, when the target has only a point scatterer, the wideband ambiguity function (WBAF), which is the counterpart of the AF, is employed \cite{Weiss1994,Jin1995,Sibul1981}. It is shown in \cite{Jin1995} that under high SNRs, the WBAF estimator is asymptotically unbiased and the variances are close to the CRLBs for a large variety of signals.  In this section, we examine the performance of the WBAF-based estimator for an extended target.

The WBAF, suggested by \cite{Jin1995}, is
\begin{equation}
W_{s_rs_d}(\tau,\gamma)=\sqrt{\gamma}\int\limits_{-\infty}^{+\infty}s_r(t)s_d^{*}(\gamma(t-\tau))dt,
\end{equation}
where $s_r$ and $s_d$ are the received and the reference signal, respectively.
The received signal $s_r$ is modeled as (\ref{equ:a22}), and the reference signal $s_d$ is different for various estimators: 1) \emph{Oracle matched filter}
$\left[\hat{\tau}_*,\hat{\gamma}_*\right]=\mathrm{arg} \max\limits_{\tau,\gamma} W_{s_rs_d}$ with $s_d=\sum\limits_{p=1}^P x_p s(\gamma(t-\tau_p))$, 2) \emph{WBAF estimator}
$\left[\hat{\tau},\hat{\gamma}\right]=\mathrm{arg} \max\limits_{\tau,\gamma} W_{s_rs_d}$ with $s_d=s(\gamma(t-\tau_p))$.
The estimates $\left[\hat{\tau}_*, \hat{\gamma}_*\right]$ are ideal but impractical, because the number of scatterers $P$ and the scattering coefficients ${\bf x}$ are unknown.  The oracle matched filter is employed as a reference to illustrate the properties of CRLBs. In practice, the WBAF estimator $\left[\hat{\tau},\hat{\gamma}\right]$ is often applied.

\begin{figure}[!t]
\centering
\includegraphics[width=3in]{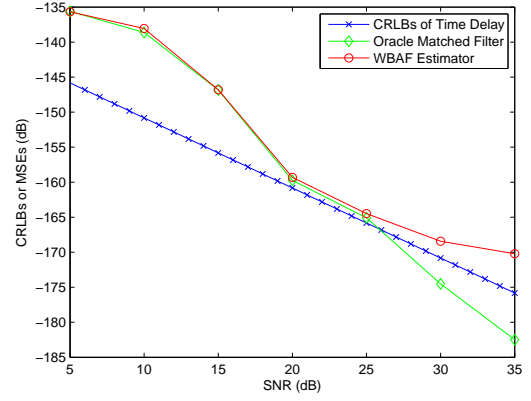}
\caption{The CRLBs and MSEs of time delay with $P=4$}
\label{fig:a17}
\end{figure}

\begin{figure}[!t]
\centering
\includegraphics[width=3in]{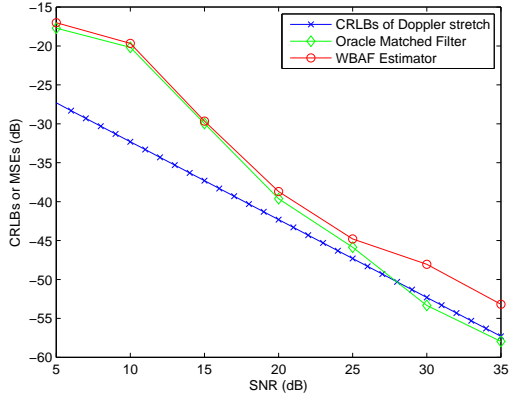}
\caption{The CRLBs and MSEs of Doppler stretch with $P=16$}
\label{fig:a18}
\end{figure}

The CRLBs and the mean square errors (MSEs) of these two estimators versus various SNRs are shown in Fig.\ref{fig:a17}-\ref{fig:a18}. The number of scatterers are $P=4$ and $16$. All the $x_p$ are assumed to equal $1$. The time delay is $\tau=2 \times 10^{-4}\mathrm{s}$ and the Doppler stretch is $\gamma=1/1.06 $. The source signal $s(t)$ is a monopulse Chirp signal, time-limited to $[0,5\times10^{-5}\mathrm{s}]$ and approximately band-limited to $1.28\times10^{5}\mathrm{Hz}$, that is,
\begin{equation}
s(t)=\mathrm{cos}(2\pi a t^2)[u(t)-u(t-T)],
\end{equation}
where $a=2.56\times 10^{9} \mathrm{Hz}/\mathrm{s}$, $T=5\times 10^{-5}\mathrm{s}$ and $u(t)$ is the unit step function. The SNR is defined as
\begin{equation}
\mathrm{SNR}=\frac{1}{N_0}\int\limits_{-\infty}^{+\infty}\left|\sum\limits_{p=1}^P x_p s(\gamma(t-\tau_p))\right|^2dt=\frac{1}{\gamma N_0}{\bf x}^T{\bf \Lambda}{\bf x}
\end{equation}
and is changed by altering $N_0$. The sampling interval $\Delta=6.25\times10^{-8}\mathrm{s}$. The CRLBs are calculated by (\ref{equ:a37})-(\ref{equ:a27}). The MSEs are computed with $100$ independent Monte Carlo trials.
As presented in Fig.\ref{fig:a17}-\ref{fig:a18}, the MSEs of the Oracle matched filter estimator are smaller than the corresponding CRLBs when the SNR is relatively large (e.g. larger than $26\mathrm{dB}$ when $P=4$) and the reason is that the Oracle matched filter assumes that all $x_p$ are known and thus the number of unknown parameters is actually smaller than the number of unknowns in the CRLB derivation. Meanwhile, the MSEs of WBAF estimator gradually deviate from the corresponding CRLBs, indicating that the WBAF estimator is not appropriate under high SNRs. In addition, we find that under high SNRs, the performance of the WBAF is significantly affected by the number of scatterers.
\begin{figure}[!t]
\centering
\includegraphics[width=3in]{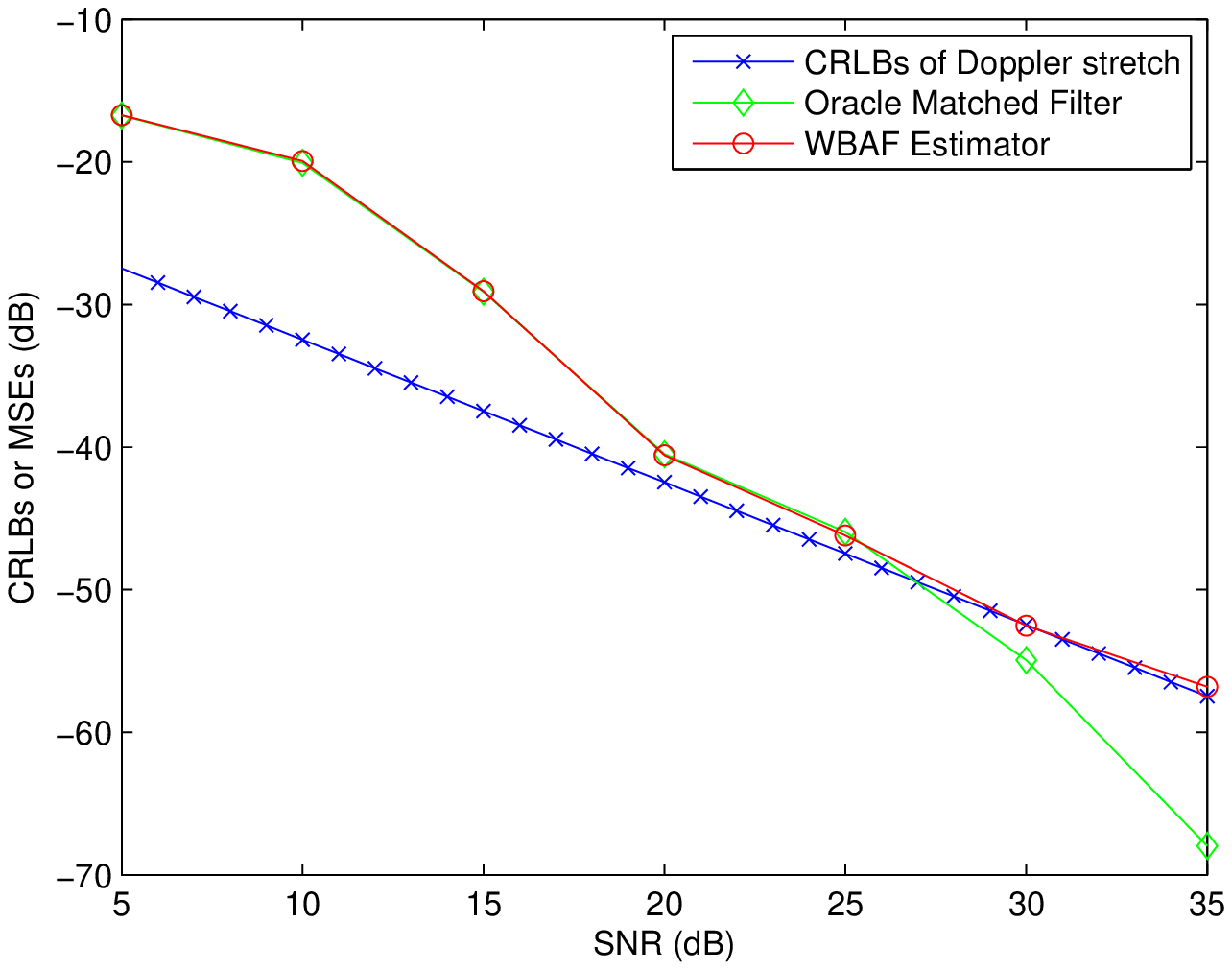}
\caption{The CRLBs and MSEs of Doppler stretch with $P=4$}
\end{figure}
\begin{figure}[!t]
\centering
\includegraphics[width=3in]{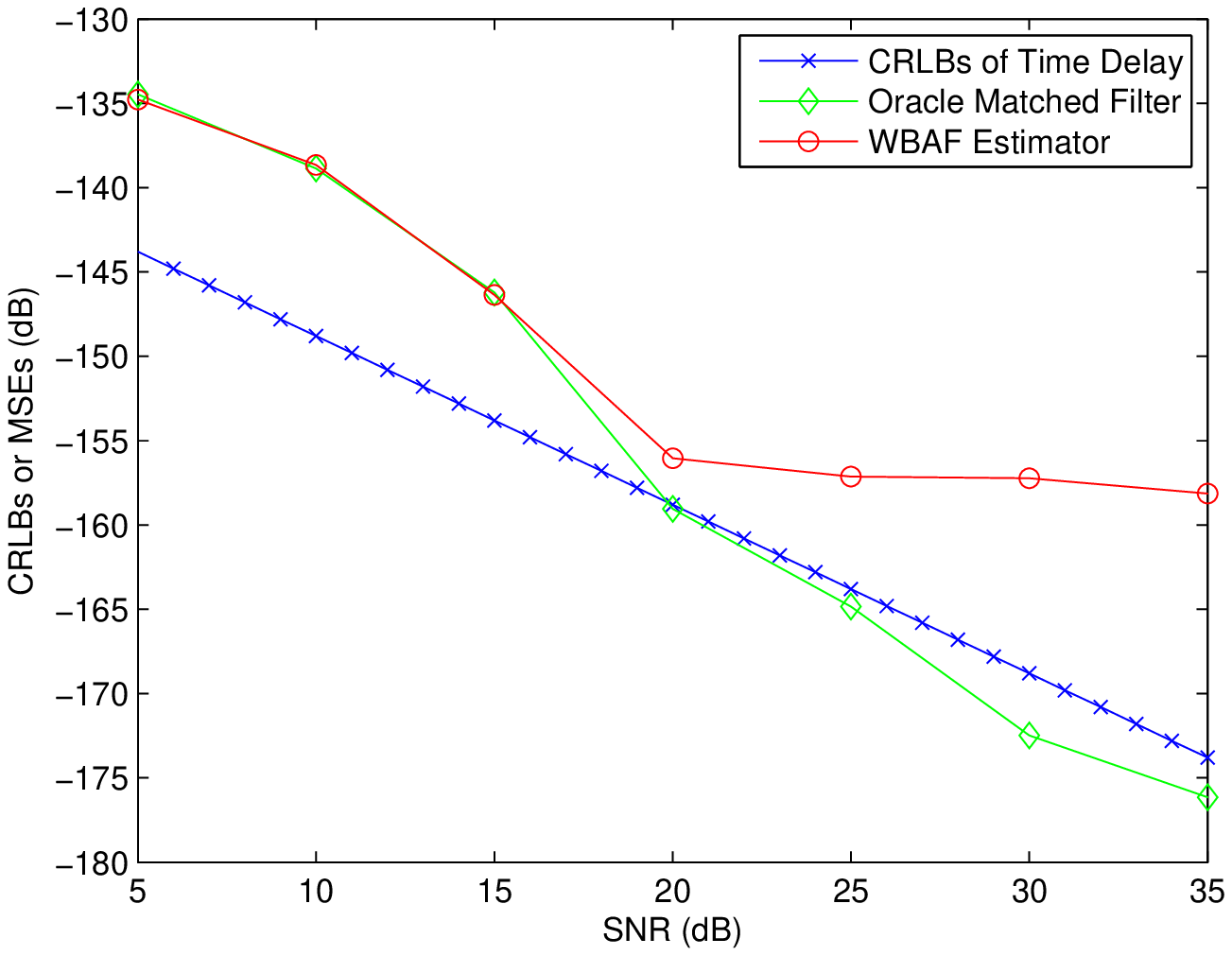}
\caption{The CRLBs and MSEs of time delay with $P=16$}
\end{figure}

\begin{figure}[!t]
\centering
\includegraphics[width=3in]{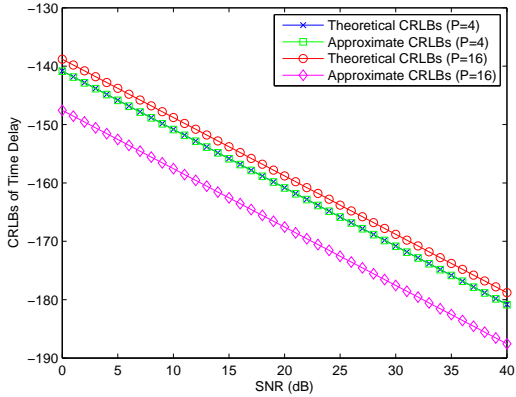}
\caption{The comparison between the theoretical and approximate CRLBs}
\label{fig:a19}
\end{figure}

\begin{figure}[!t]
\centering
\includegraphics[width=3in]{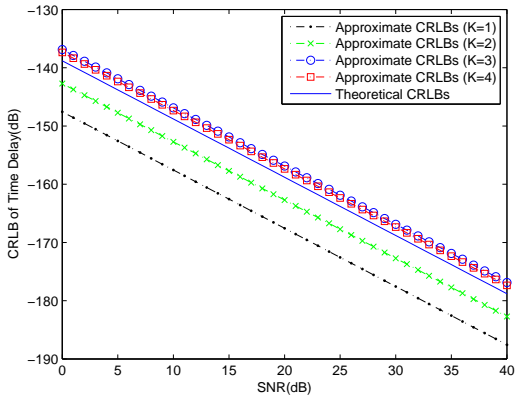}
\caption{The approximate CRLBs for different K with $P=16$ }
\label{fig:a20}
\end{figure}

The approximate CRLBs (\ref{equ:a30})-(\ref{equ:a21}) are next compared with the theoretical CRLBs (\ref{equ:a24})-(\ref{equ:a25}). The results  are presented in Fig.\ref{fig:a19} with $P=4$ and $16$, respectively. The approximate CRLBs are calculated using (\ref{equ:a30})-(\ref{equ:a21}) with $K=1$.  Other parameters are the same as those for Fig.\ref{fig:a17}. It is seen that the approximate CRLBs are accurate in the case of a small target ($P=4$) and become less accurate when the target is relatively large ($P=16$). The approximate CRLBs with $1 \le K \le4$ for $P=16$ are presented in Fig.\ref{fig:a20}. Fig.\ref{fig:a19}-\ref{fig:a20} indicate that 1) the approximate error diminishes if a larger $K$ is chosen, 2) a larger $K$ is required as the size of target increases. These statements are coincident with (\ref{equ:a83})-(\ref{equ:a84}).

\begin{figure}[!t]
\centering
\includegraphics[width=3in]{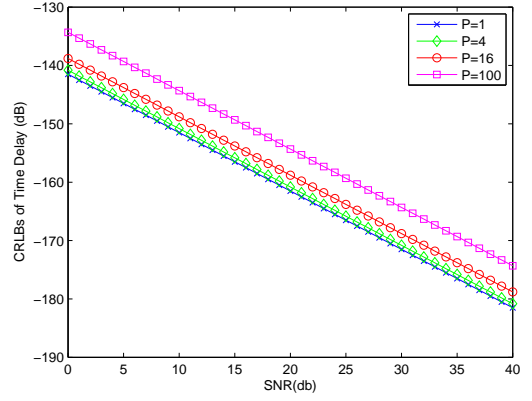}
\caption{The effects of $P$ on the CRLBs of time delay}
\label{fig:a15}
\end{figure}

\begin{figure}[!t]
\centering
\includegraphics[width=3in]{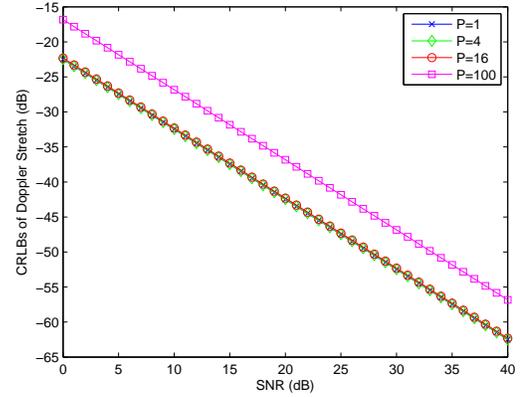}
\caption{The effects of $P$ on the CRLBs of Doppler stretch}
\label{fig:a16}
\end{figure}

The influences of the size of the target on the CRLBs are shown in Fig.\ref{fig:a15} and Fig.\ref{fig:a16}, where $P=1, 4, 16$ and $100$. The other parameters are the same as those for Fig.\ref{fig:a17}. The CRLBs are calculated with (\ref{equ:a37})-(\ref{equ:a27}). It indicates that the CRLBs are higher  when the size of target increases.

\begin{figure}[!t]
\centering
\includegraphics[width=3in]{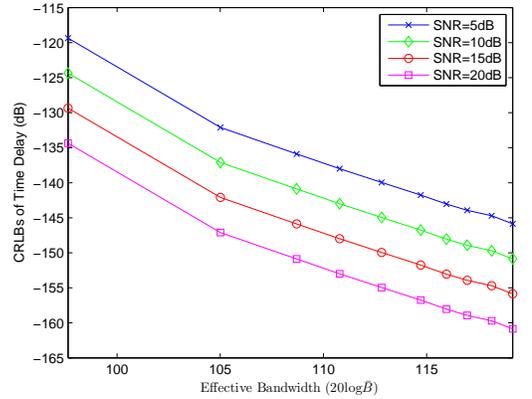}
\caption{The influences of effective bandwidth on the CRLBs of time delay}
\label{fig:a11}
\end{figure}

The influences of the effective bandwidth on the CRLBs of the time delay are shown in Fig.\ref{fig:a11}, where $a$ changes from $0.256\times 10^{9} \mathrm{Hz}/\mathrm{s}$ to $2.560\times 10^{9} \mathrm{Hz}/\mathrm{s}$ and other parameters are the same as those for Fig.\ref{fig:a17}. The effective bandwidth $\bar{B}$ increases from $0.7604\times 10^5\mathrm{Hz}$ to $9.0884\times 10^5 \mathrm{Hz}$. The effective duration $\bar{T}$  increases from $3.549\times 10^{-5}\mathrm{s}$ to $3.893\times 10^{-5} \mathrm{s}$ and can be considered as almost unchanged.  The CRLBs are calculated with (\ref{equ:a37})-(\ref{equ:a27}). These numerical results demonstrate that the CRLB of the time delay is inversely proportional to the effective bandwidth of the transmitted signal.

\begin{figure}[!t]
\centering
\includegraphics[width=3in]{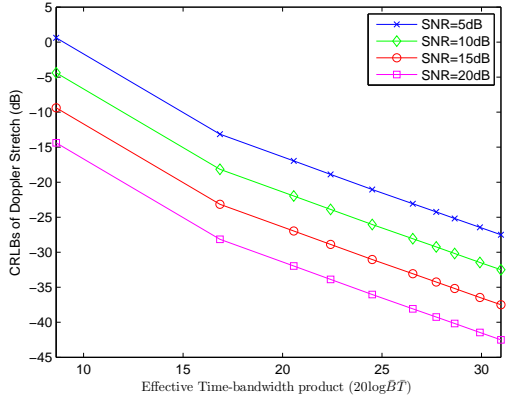}
\caption{The effects of effective time-bandwidth product on the CRLBs of Doppler stretch. $\bar{T}=(3.7\pm 0.2) \times 10^{-5}\mathrm{s}$ and is almost unchanged. }
\label{fig:a13}
\end{figure}

\begin{figure}[!t]
\centering
\includegraphics[width=3in]{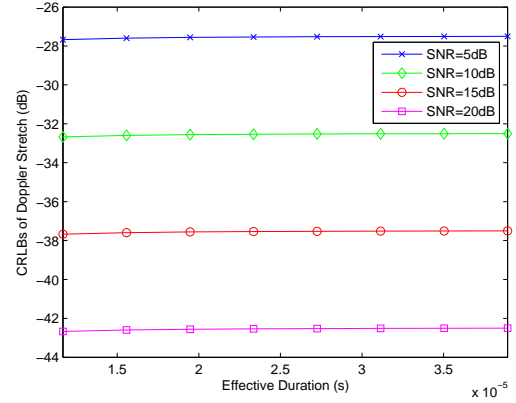}
\caption{The effects of effective time-bandwidth product on the CRLBs of Doppler stretch. $\bar{B}\bar{T}\equiv35.3786$.}
\label{fig:a14}
\end{figure}

Two experiments are performed to demonstrate the relation between the time-bandwidth product and the CRLB of the Doppler stretch. In the first one,  $\bar{B}\bar{T}$ changes and $\bar{T}$ is fixed. In the second one, $\bar{B}\bar{T}$ is fixed and $\bar{T}$ varies. The results are depicted in Fig.\ref{fig:a13} and Fig.\ref{fig:a14}, respectively. Note that the effective time-bandwidth product $\bar{B}\bar{T}$ is proportional to $aT^2$ for a Chirp signal. In Fig.\ref{fig:a13}, $a$ changes from $0.256\times 10^{9} \mathrm{Hz}/\mathrm{s}$ to $2.560\times 10^{9} \mathrm{Hz}/\mathrm{s}$ and other parameters are the same as those for Fig.\ref{fig:a17}. The effective time-bandwidth product $\bar{B}\bar{T}$ increases from $2.6988$ to $35.3786$. The effective duration $\bar{T}$ increases from $3.549\times 10^{-5}\mathrm{s}$ to $3.893\times 10^{-5} \mathrm{s}$ and can be considered as almost unchanged. These parameters are designed similarly to those for Fig.\ref{fig:a11}. In Fig.\ref{fig:a14}, $aT^2\equiv6.4$, $T$ increases from $1.5 \times 10^{-5}\mathrm{s}$ to $5\times 10^{-5} \mathrm{s}$ and other parameters are the same as those for Fig.\ref{fig:a17}, implying that $\bar{T}$ increases from $1.1678\times 10^{-5}\mathrm{s}$ to $3.8927\times 10^{-5} \mathrm{s}$ and $\bar{B}\bar{T}\equiv35.3786$. The CRLBs in both figures are calculated with (\ref{equ:a37})-(\ref{equ:a27}).  Combining Fig.\ref{fig:a13} with Fig.\ref{fig:a14}, we find 1) there exists a positive correlation between the effective time-bandwidth product and the estimation accuracy of the Doppler stretch, 2) the relation between the effective duration and the CRLB of the Doppler stretch is not apparent.

\section{Conclusion}{\label{sec:a7}}
In this paper, both integral and series representations of the CRLBs for the joint delay-Doppler estimation of an extended target are derived. Based on series expansion, approximations
of CRLBs are obtained. Our theoretical analyses and numerical examples indicate that the CRLBs of the time delay and the Doppler stretch are inversely proportional to the effective bandwidth and the effective time-bandwidth product, respectively. In addition, compared with the case involving a single scatterer, an extended target consisting of multiple scatterers leads to higher CRLBs under the same SNR level.


%

\appendices
\section{Theorem \ref{theorem:12} and its proof}{\label{app:05}}
\begin{theorem}{\label{theorem:12}}
For $p,q \in \mathbb{N}^{+}$, we have
\begin{equation}{\label{equ:a38}}
\begin{split}
\mathrm{Re}& \left\{ \int_{-\infty}^{+\infty}  s^{*(p)}(t)s^{(q)}(t)dt\right\}\\
 &=\left\{ \begin{array}{ll}
(-1)^{p+k} M_0^{(k)} , &p+q=2k,\\
0 , &p+q=2k+1,
\end{array}
\right.
\end{split}
\end{equation}
\begin{equation}{\label{equ:a43}}
\begin{split}
\mathrm{Re} &\left\{\int_{-\infty}^{+\infty}
ts^{*(p)}(t)s^{(q)}(t)dt\right\}\\
&= \left\{ \begin{array}{ll}
(-1)^{p+k} M_1^{(k)}, &p+q=2k,\\
(-1)^{p+k} (p-k-\frac{1}{2})M_0^{(k)},&p+q=2k+1,
\end{array}
\right.
\end{split}
\end{equation}
\begin{align}{\label{equ:a44}}
&\mathrm{Re} \left\{ \int_{-\infty}^{+\infty}
t^2s^{*}(t)s^{(q)}(t)dt\right\}\\
&= \left\{ \begin{array}{ll}
(-1)^{k}M_2^{(k)}+(-1)^{k+1} k^2M_0^{(k-1)}, &q=2k,\nonumber\\
(-1)^{k+1}(2k+1)M_1^{(k)}, &q=2k+1,\nonumber
\end{array}
\right.
\end{align}
\begin{equation}{\label{equ:a100}}
\begin{split}
\mathrm{Im}& \left\{ \int_{-\infty}^{+\infty}  s^{*(p)}(t)s^{(q)}(t)dt\right\}\\
 &=\left\{ \begin{array}{ll}
0, &p+q=2k,\\
 (-1)^{p+k} \widetilde{M}_0^{(k)}, &p+q=2k+1,
\end{array}
\right.
\end{split}
\end{equation}
\begin{equation}
\begin{split}
\mathrm{Im} &\left\{\int_{-\infty}^{+\infty}
ts^{*(p)}(t)s^{(q)}(t)dt\right\}\\
&= \left\{ \begin{array}{ll}
(-1)^{p+k} (k-p)\widetilde{M}_0^{(k-1)}, &p+q=2k,\\
(-1)^{p+k} \widetilde{M}_1^{(k)},&p+q=2k+1,
\end{array}
\right.
\end{split}
\end{equation}
\begin{align}{\label{equ:a101}}
&\mathrm{Im} \left\{ \int_{-\infty}^{+\infty}
t^2s^{*}(t)s^{(q)}(t)dt\right\}\\
&= \left\{ \begin{array}{ll}
(-1)^{k}2k\widetilde{M}_1^{(k-1)}, &q=2k,\nonumber\\
(-1)^{k}\widetilde{M}_2^{(k)}-(-1)^{k}(k^2+k)\widetilde{M}_0^{(k-1)}, &q=2k+1.\nonumber
\end{array}
\right.
\end{align}
where $^*$ denotes the complex conjugate.
\end{theorem}
\emph{Proof of (\ref{equ:a43})}.
\begin{IEEEproof}
Write $s(t)$ in the form of  $u(t)+jv(t)$. Then, for $m=2k, k \in \mathbb{N}^+$, we have
\begin{align}{\label{equ:a50}}
&\mathrm{Re}  \left\{\int_{-\infty}^{+\infty}  s^{*(0)}(t)s^{(m)}(t)dt\right\} \nonumber\\
&=\int_{-\infty}^{+\infty}  u(t)u^{(2k)}(t)+v(t)v^{(2k)}(t)dt \nonumber\\
&=(-1)\left( \int_{-\infty}^{+\infty}  u^{(1)}(t)u^{(2k-1)}(t)+v^{(1)}(t)v^{(2k-1)}(t)dt\right)\nonumber\\
&=(-1)^k\left( \int_{-\infty}^{+\infty}  u^{(k)}(t)u^{(k)}(t)+v^{(k)}(t)v^{(k)}(t)dt\right) \nonumber\\
&= (-1)^k M_0^{(k)} .
\end{align}
Similarly, for $m=2k+1, k \in \mathbb{N}$,
\begin{align}
\mathrm{Re} &\left\{\int_{-\infty}^{+\infty}  s^{*(0)}(t)s^{(m)}(t)dt\right\}=\nonumber\\
&-\mathrm{Re} \left\{\int_{-\infty}^{+\infty}  s^{*(0)}(t)s^{(m)}dt\right\},
\end{align}
which implies
\begin{align}
\mathrm{Re} \left\{\int_{-\infty}^{+\infty}  s^{*(0)}(t)s^{(2k+1)}(t)dt\right\}=0.
\end{align}
Finally, for $p,q \in \mathbb{N}$, (\ref{equ:a38}) is derived as follows
\begin{align}{\label{equ:a41}}
\mathrm{Re}& \left\{\int_{-\infty}^{+\infty}  s^{*(p)}(t)s^{(q)}(t)dt\right\} \nonumber\\
&=(-1)^p \mathrm{Re} \left\{\int_{-\infty}^{+\infty}  s^{*(0)}(t)s^{(p+q)}(t)dt\right\} \nonumber\\
&=\left\{ \begin{array}{ll}
(-1)^{p+k} M_0^{(k)} , &p+q=2k.\\
0 , &p+q=2k+1.
\end{array}
\right.
\end{align}
\end{IEEEproof}
\emph{Proof of (\ref{equ:a43})}.
\begin{IEEEproof}
For $m,n \in \mathbb{N}^+$ and $n \le m$, we have
\begin{align}{\label{equ:a42}}
\mathrm{Re} &\left\{\int_{-\infty}^{+\infty} t s^{*(0)}(t)s^{(m)}(t)dt\right\} \nonumber\\
&= (-1)\mathrm{Re}\left\{\int_{-\infty}^{+\infty} t s^{*(1)}(t)s^{(m-1)}(t)dt\right\}  \nonumber\\
&+(-1)\mathrm{Re} \left\{\int_{-\infty}^{+\infty}  s^{*(0)}(t)s^{(m-1)}(t)dt\right\}  \nonumber\\
&= (-1)^n \mathrm{Re} \left\{\int_{-\infty}^{+\infty} t s^{*(n)}(t)s^{(m-n)}(t)dt\right\}  \nonumber\\
&+\sum\limits_{l=0}^{n-1}(-1)^{l+1} \mathrm{Re} \left\{\int_{-\infty}^{+\infty}  s^{*(l)}(t)s^{(m-1-l)}(t)dt\right\}.
\end{align}
By making use of (\ref{equ:a41}), the second term in the last line of (\ref{equ:a42}) becomes
\begin{align}
\sum\limits_{l=0}^{n-1}&(-1)^{l+1} \mathrm{Re} \left\{\int_{-\infty}^{+\infty}  s^{*(l)}(t)s^{(m-1-l)}(t)dt\right\}\nonumber \\
&=\left\{ \begin{array}{ll}
(-1)^{k+1}n M_0^{(k)}, &m-1=2k.\\
0,&m-1=2k+1.
\end{array}
\right.
 \end{align}
Thus, if $m=2k$, let $n=k$, and (\ref{equ:a42}) becomes
\begin{align}{\label{equ:a53}}
\mathrm{Re} &\left\{\int_{-\infty}^{+\infty} t s^{*(0)}(t)s^{(2k)}dt\right\}= (-1)^{k}M_1^{(k)},
\end{align}
if $m=2k+1$, let $n=m$, and (\ref{equ:a42}) becomes
\begin{align}
\mathrm{Re} &\left\{\int_{-\infty}^{+\infty} t s^{*(0)}(t)s^{(2k+1)}dt\right\}=(-1)^{k+1}(2k+1)M_0^{(k)}+\nonumber\\
&(-1)^{(2k+1)}\mathrm{Re} \left\{\int_{-\infty}^{+\infty} t s^{*(2k+1)}(t)s^{(0)}dt\right\},
\end{align}
which implies
\begin{align}{\label{equ:a54}}
\mathrm{Re} \left\{\int_{-\infty}^{+\infty} t s^{*(0)}(t)s^{(2k+1)}dt\right\}=(-1)^{k+1}(k+\frac{1}{2})M_0^{(k)}.
\end{align}
With (\ref{equ:a41}) (\ref{equ:a53}) and (\ref{equ:a54}), (\ref{equ:a43}) is derived as follows
\begin{align}{\label{equ:a45}}
\mathrm{Re}& \left\{\int_{-\infty}^{+\infty} t s^{*(p)}(t)s^{(q)}(t)dt\right\} \\
&= (-1)\mathrm{Re}\left\{\int_{-\infty}^{+\infty} t s^{*(p-1)}(t)s^{(q+1)}(t)dt\right\}  \nonumber\\
&+(-1)^{p}\mathrm{Re} \left\{\int_{-\infty}^{+\infty}  s^{*(0)}(t)s^{(p+q-1)}(t)dt\right\}  \nonumber\\
&=(-1)^p \mathrm{Re} \left\{\int_{-\infty}^{+\infty} t s^{*(0)}(t)s^{(p+q)}dt\right\} \nonumber\\
& +(-1)^{p}p \mathrm{Re} \left\{\int_{-\infty}^{+\infty}  s^{*(0)}(t)s^{(p+q-1)}dt\right\}  \nonumber\\
&=\left\{ \begin{array}{ll}
(-1)^{p+k} M_1^{(k)}, &p+q=2k.\\
(-1)^{p+k} (p-k-\frac{1}{2})M_0^{(k)},&p+q=2k+1.
\end{array}\nonumber
\right.
\end{align}
\end{IEEEproof}
\emph{Proof of (\ref{equ:a44})}.
\begin{IEEEproof}
For $m,n \in \mathbb{N}^+$ and $n \le m$, we have
\begin{align} {\label{equ:a46}}
\mathrm{Re} &\left\{\int_{-\infty}^{+\infty} t^2 s^{*(0)}(t)s^{(m)}(t)dt\right\} \\
&= (-1)\mathrm{Re} \left\{\int_{-\infty}^{+\infty} t^2 s^{*(1)}(t)s^{(m-1)}(t)dt\right\} \nonumber\\
& \ \ \ +(-2)\mathrm{Re} \left\{\int_{-\infty}^{+\infty} t s^{*(0)}(t)s^{(m-1)}(t)dt\right\} \nonumber\\
&=(-1)^{n}\mathrm{Re} \left\{\int_{-\infty}^{+\infty} t^2 s^{*(n)}(t)s^{(m-n)}(t)dt\right\}\nonumber\\
&+\sum\limits_{l=0}^{n-1}(-1)^{l+1} 2\mathrm{Re} \left\{\int_{-\infty}^{+\infty} t s^{*(l)}(t)s^{(m-1-l)}(t)dt\right\}.  \nonumber
\end{align}
By making use of (\ref{equ:a45}), the second term in the last line of (\ref{equ:a46}) becomes
\begin{align}
&\sum\limits_{l=0}^{n-1}(-1)^{l+1} 2\mathrm{Re} \left\{\int_{-\infty}^{+\infty} t s^{*(l)}(t)s^{(m-1-l)}(t)dt\right\} \\
&=\left\{ \begin{array}{ll}
(-1)^{k+1}2n M_1^{(k)}, &m-1=2k.\\
(-1)^{k+1}(n^2-2(k+1)n)M_0^{(k)},&m-1=2k+1.
\end{array}
\right. \nonumber
\end{align}
Thus, if $m=2k$, let $n=k$, and (\ref{equ:a46}) becomes
\begin{align} {\label{equ:a47}}
\mathrm{Re}& \left\{\int_{-\infty}^{+\infty} t^2 s^{*(0)}(t)s^{(m)}(t)dt\right\}\nonumber\\
&=(-1)^{k}M_2^{(k)}+(-1)^{k+1}k^2M_0^{k-1},
\end{align}
if $m=2k+1$, let $n=m$, and (\ref{equ:a46}) becomes
\begin{align}
\mathrm{Re} &\left\{\int_{-\infty}^{+\infty} t^2 s^{*(0)}(t)s^{(m)}(t)dt\right\}=(-1)^{k+1}2(2k+1)M_1^{(k)}+\nonumber\\
&(-1)^{2k+1}\mathrm{Re} \left\{\int_{-\infty}^{+\infty} t^2 s^{*(m)}(t)s^{(0)}(t)dt\right\},
\end{align}
which implies
\begin{align}{\label{equ:a48}}
\mathrm{Re} &\left\{\int_{-\infty}^{+\infty} t^2 s^{*(0)}(t)s^{(m)}(t)dt\right\}=(-1)^{k+1}(2k+1)M_1^{(k)}.
\end{align}
Combining (\ref{equ:a47}) and (\ref{equ:a48}) gives (\ref{equ:a44}).
\end{IEEEproof}
The proofs of (\ref{equ:a100})-(\ref{equ:a101}) are similar and thus are omitted.

\section{The series representations of the CRLBs}{\label{app:a13}}
The CRLBs in the form of series are given by (\ref{equ:a24}) (\ref{equ:a25}) and (\ref{equ:a87}), where
\begin{align}{\label{equ:a28}}
F_{ij}=\lim\limits_{K \to \infty}F_{ij}^{(K)}=\lim\limits_{K \to \infty}F_{1ij}^{(K)}+\sqrt{-1}F_{2ij}^{(K)},
\end{align}
\begin{align}{\label{equ:a32}}
\overline{{\bf F}}_{3i}=\lim\limits_{K\rightarrow +\infty} \overline{{\bf F}}_{3i}^{(K)}=\lim\limits_{K\rightarrow +\infty} {\bf F}_{13i}^{(K)}+\sqrt{-1} {\bf F}_{23i}^{(K)},
\end{align}
and
\begin{equation}\label{equ:a39}
F_{111}^{(K)} =\sum\limits_{0 \le 2k \le K}   \frac{(-1)^k 2\gamma^{2k+1}  }{
(2k)!N_0 }M_0^{(k+1)}
{\bf x}^H{\bf \Gamma }^{(2k)}{\bf x},
\end{equation}
\begin{equation}
F_{211}^{(K)}=\sum\limits_{0 \le 2k+1 \le K}   \frac{(-1)^{k} 2\gamma^{2k+2}  }{
(2k+1)!N_0 }\widetilde{M}_0^{(k+1)}
{\bf x}^H{\bf \Gamma }^{(2k+1)}{\bf x},
\end{equation}
\begin{equation}
F_{112}^{(K)}=\sum\limits_{0 \le 2k \le K}  \frac{(-1)^{k+1}2\gamma^{2k-1}}{(2k)!N_0  }  M_1^{(k+1)}{\bf x}^H{\bf \Gamma }^{(2k)}{\bf x}, \\
\end{equation}
\begin{equation}
F_{212}^{(K)}=\sum\limits_{0 \le 2k+1 \le K}  \frac{(-1)^{k+1}2\gamma^{2k}}{(2k+1)!N_0  }  \widetilde{M}_1^{(k+1)}{\bf x}^H{\bf \Gamma }^{(2k+1)}{\bf x}, \\
\end{equation}
\begin{align}
F_{122}^{(K)} &=\sum\limits_{1 \le 2k \le K}   \frac{(-1)^{k}(k-1) \gamma^{2k-3} }{(2k-1)! N_0 } M_0^{(k)} {\bf x}^H{\bf \Gamma }^{(2k)}{\bf x} +  \nonumber \\
&\ \  \sum\limits_{0 \le 2k \le K}  \frac{(-1)^{k}2 \gamma^{2k-3}}{(2k)!N_0 }  M_2^{(k+1)}{\bf x}^H{\bf \Gamma }^{(2k)}{\bf x},
\end{align}
\begin{align}
F_{222}^{(K)} &=\sum\limits_{0 \le 2k+1 \le K}   \frac{(-1)^{k}2k^2\gamma^{2k-2} }{(2k+1)! N_0 } \widetilde{M}_0^{(k)} {\bf x}^T{\bf \Gamma }^{(2k+1)}{\bf x} +  \nonumber \\
&\ \  \sum\limits_{0 \le 2k+1 \le K}  \frac{(-1)^{k}2 \gamma^{2k-2}}{(2k+1)!N_0 } \widetilde{ M}_2^{(k+1)}{\bf x}^H{\bf \Gamma }^{(2k+1)}{\bf x},
\end{align}
\begin{align}
{\bf F}^{(K)}_{131}&=\sum\limits_{0 \le 2k-1 \le K}\frac{(-1)^{k+1}
2\gamma^{2k-1}}{(2k-1)!N_0}M_0^{(k)}{\bf \Gamma }^{(2k-1)}{\bf x},
\end{align}
\begin{align}
{\bf F}^{(K)}_{231}&=\sum\limits_{0 \le 2k \le K}\frac{(-1)^{k+1}
2\gamma^{2k}}{(2k)!N_0}\widetilde{M}_0^{(k)}{\bf \Gamma }^{(2k)}{\bf x},
\end{align}
\begin{align}
{\bf F}^{(K)}_{132}&= \sum\limits_{0 \le 2k \le K} \frac{(-1)^{k}(2k-1) \gamma^{2k-2} }{(2k)!N_0 } M_0^{(k)} {\bf \Gamma }^{(2k)}{\bf x}+  \nonumber \\
&\ \   \sum\limits_{0 \le 2k+1 \le K} \frac{(-1)^{k+1}2 \gamma^{2k-1} }{(2k+1)! N_0} M_1^{(k+1)} {\bf \Gamma }^{(2k+1)}{\bf x},
\end{align}
\begin{align}
{\bf F}^{(K)}_{232}&= \sum\limits_{0 \le 2k+1 \le K} \frac{(-1)^{k}2k\gamma^{2k-1} }{(2k+1)!N_0 } \widetilde{M}_0^{(k)} {\bf \Gamma }^{(2k+1)}{\bf x}+  \nonumber \\
&\ \  \sum\limits_{0 \le 2k \le K} \frac{(-1)^{k}2 \gamma^{2k-2} }{(2k)! N_0} \widetilde{M}_1^{(k)} {\bf \Gamma }^{(2k)}{\bf x},
\end{align}
\begin{align}
{\bf F}^{(K)}_{133}&= \sum\limits_{0 \le 2k \le K}\frac{(-1)^{k} 2\gamma^{2k-1}}{(2k)!
N_0 } M_0^{(k)} {\bf \Gamma }^{(2k)},
\end{align}
\begin{align}{\label{equ:a29}}
{\bf F}^{(K)}_{233}&= \sum\limits_{0 \le 2k+1 \le K}\frac{(-1)^{k} 2\gamma^{2k}}{(2k+1)!
N_0 } \widetilde{M}_0^{(k)} {\bf \Gamma }^{(2k+1)}.
\end{align}

\section{Proof of Proposition \ref{theorem:13}}\label{app:a15}
Note that
$\Gamma^{(k)}_{ij}=O((2L/c)^k)$, $M_i^{(k)}=O\left(e^{kC_2}\right)$, $\widetilde{M}_i^{(k)}=O\left(e^{kC_2}\right)$.
Thus, from (\ref{equ:a28})-(\ref{equ:a29}), we obtain
\begin{equation}
F_{ij}-F_{ij}^{(K)}=O\left(\frac{ \left( 2L \gamma \exp\{C_2\}/ c\right)^{K+1}}{(K+1)!}\right),
\end{equation}
\begin{equation}
\overline{{\bf F}}_{3i}-\overline{{\bf F}}_{3i}^{(K)}=O\left(\frac{ \left( 2L \gamma \exp\{C_2\}/ c\right)^{K+1}}{(K+1)!}\right).
\end{equation}
Because ${\bf E}^{(K)}=\overline{{\bf F}}_{33}-\overline{{\bf F}}_{33}^{(K)} \rightarrow 0$ as $K \rightarrow +\infty$, the inverse of $\overline{{\bf F}}_{33}$ can be written as \cite{Zorich2004II}
\begin{align}{\label{equ:a49}}
\overline{{\bf F}}_{33} ^{-1}=\left(\overline{{\bf F}}_{33}^{(K)}\right)^{-1}+O\left(
{\bf E}^{(K)} \right), K\rightarrow+\infty,
\end{align}
where
\begin{equation}{\label{equ:a31}}
 O\left({\bf E}^{(K)} \right)=\sum\limits_{n=1}^{+\infty}\left(-\left(\overline{{\bf F}}_{33}^{(K)}\right)^{-1}
{\bf E}^{(K)}\right)^n \left(\overline{{\bf F}}_{33}^{(K)}\right)^{-1}.
\end{equation}
Thus, we have
\begin{align}
&\left\| \left(\overline{{\bf F}}_{33}^{(K)} \right)^{-1}
-\overline{{\bf F}}_{33}^{-1}\right\| \le \frac{\left\|\left(\overline{{\bf F}}_{33}^{(K)}\right)^{-1}
\right\|^2\left\|{\bf E}^{(K)}\right\|}{1-\left\|\left(\overline{{\bf F}}_{33}^{(K)}\right)^{-1}
{\bf E}^{(K)}\right\|} \nonumber\\
&=O\left(\frac{ \left( 2L \gamma \exp\{C_2\}/ c\right)^{K+1}}{(K+1)!}\right),
\end{align}
and (\ref{equ:a83})-(\ref{equ:a84}) follow.
\section{Proof of Theorem \ref{theorem:a13} and \ref{theorem:11} }\label{app:a12}
\begin{lemma}
Let $M_0\le +\infty$,  $B_0\le +\infty$ and $T_0<+\infty$. Assume that there exists a constant $\epsilon>0$ such that
\begin{equation}
\lim\limits_{ \left(M_0^{(0)}, \bar{B}, \bar{T}\right) \rightarrow \left(M_0,B_0,T_0\right) }m\left(\left\{t \left|\left|s^{(1)}(t)\right|>\epsilon \right. \right\}\right)>0.
\end{equation}
Then, there exists a constant $C_5 \in (0,1)$ such that
\begin{equation}{\label{equ:a72}}
\bar{T}^2/T^2<\left(M_1^{(1)} \right)^2/ M_0^{(1)}M_2^{(1)} \leq C_5,
\end{equation}
as $\left(M_0^{(0)}, \bar{B}, \bar{T}\right) \rightarrow \left(M_0,B_0,T_0\right)$.
\end{lemma}
\begin{IEEEproof}
According to the Cauchy-Schwartz inequality \cite{H.L.Royden1988} and $s(t)=0, t \notin[0,T]$, we have
\begin{equation}
\left(M_2^{(1)}/T\right)^2< \left(M_1^{(1)} \right)^2 \le M_0^{(1)}M_2^{(1)},
\end{equation}
which implies
\begin{equation}{\label{equ:a63}}
\bar{T}^2/T^2<\left(M_1^{(1)} \right)^2/M_0^{(1)}M_2^{(1)} \le 1.
\end{equation}
Suppose ({\ref{equ:a72}}) does not. Then we have
\begin{equation}
\limsup_{\left(M_0^{(0)}, \bar{B}, \bar{T}\right) \rightarrow \left(M_0,B_0,T_0\right)} \left(M_1^{(1)} \right)^2-M_0^{(1)}M_2^{(1)}=0.
\end{equation}
Define $<f(t),g(t)>=\int_{-\infty}^{+\infty} f^{*}(t)g(t)dt$ and $||f||=<f,f>^{\frac{1}{2}}$. Thus, let $\left(M_0^{(0)}, \bar{B}, \bar{T}\right) \rightarrow \left(M_0,B_0,T_0\right)$, and we have
\begin{align}
&\left\|s^{(1)}(t)t-T_0s^{(1)}(t)\right\|^2-\left(\left\|s^{(1)}(t)t\right\|-T_0\left\|s^{(1)}(t)\right\|\right)^2 \nonumber\\
&= -2T_0\left<s^{(1)}(t)t,s^{(1)}(t)\right>+ 2T_0\left\|s^{(1)}(t)t\right\|\left\|s^{(1)}(t)\right\|\nonumber\\
&\rightarrow 0.
\end{align}
Therefore, we obtain $\left\|s^{(1)}(t)t-T_0s^{(1)}(t)\right\|^2 \rightarrow 0$, which implies $\left|s^{(1)}(t)t-T_0s^{(1)}(t)\right|\rightarrow 0$ a.e., and thus $\left|s^{(1)}(t)\right| \rightarrow 0$ a.e., as  $\left(M_0^{(0)}, \bar{B}, \bar{T}\right) \rightarrow \left(M_0,B_0,T_0\right)$. The lemma follows by contradiction.
\end{IEEEproof}
\emph{Proof of Theorem \ref{theorem:a13}}
\begin{IEEEproof}
Substituting (\ref{equ:a62})-(\ref{equ:a57}) into (\ref{equ:a56})-(\ref{equ:a69}), we have
\begin{align}{\label{equ:a103}}
&\mathrm{CRLB}_{\tau, P=1}=\nonumber\\
& \frac{N_0}{2\gamma x^2}\frac{\left(M_0^{(0)}\right)^{-1} \bar{B}^{-2}\left(1-\frac{1}{4}\bar{B}^{-2}\bar{T}^{-2}\right)}{1-\frac{1}{4}\bar{B}^{-2}\bar{T}^{-2}-\left(M_1^{(1)}\right)^2\left(M_0^{(0)}\right)^{-2}\bar{B}^{-4}\bar{T}^{-2}}\nonumber\\
&=O\left( \left(M_0^{(0)}\right)^{-1} \bar{B}^{-2}\right),
\end{align}
\begin{align}{\label{equ:a104}}
&\mathrm{CRLB}_{\gamma, P=1}=\nonumber\\
&\frac{\gamma x^3 N_0}{2x^2}\frac{\left(M_0^{(0)}\right)^{-1}\bar{B}^{-2}\bar{T}^{-2}}{1-\frac{1}{4}\bar{B}^{-2}\bar{T}^{-2}-\left(M_1^{(1)}\right)^2\left(M_0^{(0)}\right)^{-2}\bar{B}^{-4}\bar{T}^{-2}
}\nonumber\\
&=O\left(\left(M_0^{(0)}\right)^{-1}\bar{B}^{-2}\bar{T}^{-2}\right),
\end{align}
as $\left(M_0^{(0)}, \bar{B}, \bar{T} \right)\rightarrow \left(M_0,B_0,T_0\right)$. Notice that the denominators of (\ref{equ:a103}) and (\ref{equ:a104}) do not converge to zero due to (\ref{equ:a72}) and the positive definite property of ${\bf FIM}$.
\end{IEEEproof}
\emph{Proof of Theorem \ref{theorem:11}}
\begin{IEEEproof}
Referring to (\ref{equ:a28})-(\ref{equ:a29}) and (\ref{equ:a63}), we have
\begin{align}{\label{equ:a60}}
&F_{11}=O\left(M_0^{(0)} \bar{B}^2\right),F_{12}=O\left(M_0^{(0)}\bar{B}^2\bar{T}\right),\nonumber\\
&F_{22}=O\left(M_0^{(0)} \bar{B}^2\bar{T}^2\right),{\bf F}_{31}=O\left(M_0^{(0)} \bar{B}^2 {\bf \Gamma}^{(1)}{\bf x} \right),\nonumber\\
&{\bf F}_{32}=O\left(M_0^{(0)} \bar{B}^2\bar{T}{\bf \Gamma }^{(1)}{\bf x}\right),{\bf F}_{33}=O\left(M_0^{(0)} \bar{B}^2\Lambda \right).
\end{align}
Substitute (\ref{equ:a60}) into (\ref{equ:a106}), and we have
\begin{align}{\label{equ:a70}}
a_{11}&= O\left(M_0^{(0)}\bar{B}^2\right)-O\left(M_0^{(0)}\bar{B}^2 {\bf x}^T \left({\bf \Gamma }^{(1)}\right)^T \right)\times \nonumber\\
&O\left(\frac{1}{M_0^{(0)}\bar{B}^2}{\bf \Lambda }^{-1}\right)O\left(M_0^{(0)}\bar{B}^2 {\bf \Gamma }^{(1)}{\bf x} \right)\nonumber\\
&=O\left(M_0^{(0)}\bar{B}^2\right)+O\left(M_0^{(0)}\bar{B}^2 {\bf x}^T \left({\bf \Gamma }^{(1)}\right)^T {\bf \Lambda }^{-1} {\bf \Gamma }^{(1)} {\bf x} \right)\nonumber \\
&=O\left(M_0^{(0)}\bar{B}^2\right),
\end{align}
and
\begin{equation}{\label{equ:a71}}
a_{12}=O\left(M_0^{(0)}\bar{B}^2\bar{T}\right),a_{22}\approx O\left(M_0^{(0)}\bar{B}^2\bar{T}^2\right).
\end{equation}
Then, (\ref{equ:a73})-(\ref{equ:a74}) follow by substituting (\ref{equ:a70})-(\ref{equ:a71}) into (\ref{equ:a24})-(\ref{equ:a25}). Notice that the denominators of (\ref{equ:a24}) and (\ref{equ:a25}) do not converge to zero due to the positive definite property of ${\bf FIM}$.
\end{IEEEproof}


\section*{Acknowledgment}
The authors would like to thank Prof. Hongbin Li, Hongyu Gu, Wei Rao and Chu Pi for their insightful comments and suggestions.

\ifCLASSOPTIONcaptionsoff
  \newpage
\fi



\bibliographystyle{IEEEtran}
\bibliography{reference}
\end{document}